\documentclass[aps,showpacs,pre,twocolumn,superscriptaddress,floatfix]{revtex4-1}
\usepackage{graphicx,bm,amssymb,amsmath,dcolumn,hyperref}
\usepackage{multirow}
\usepackage{color}
\usepackage{subfigure}  
\usepackage{epstopdf}
\usepackage{bbm}
\usepackage{graphicx}
\usepackage{hyperref}
\usepackage{threeparttable}
\usepackage{amsthm,enumitem,mathrsfs}
\usepackage[marginal]{footmisc}
\usepackage{array}
\usepackage{tabularx}
\usepackage{makecell}

\newif\ifshowedit
\showedittrue   
\showeditfalse 

\ifshowedit
  \newcommand{\edit}[1]{{\color{blue}#1}}
  
\else
  \newcommand{\edit}[1]{#1}
  
\fi

\begin{document}
	
\title{Scaling of Long-Range Loop-Erased Random Walks}
	
\author{Tianning Xiao}
\thanks{These two authors contributed equally to this work}
\affiliation{Hefei National Research Center for Physical Sciences at the Microscale and School of Physical Sciences, University of Science and Technology of China, Hefei 230026, China}

\author{Xianzhi Pan}
\thanks{These two authors contributed equally to this work}
\affiliation{Department of Modern Physics, University of Science and Technology of China, Hefei, Anhui 230026, China}

\author{Zhijie Fan}
\email{zfanac@ustc.edu.cn}
\affiliation{Hefei National Research Center for Physical Sciences at the Microscale and School of Physical Sciences, University of Science and Technology of China, Hefei 230026, China}
\affiliation{Hefei National Laboratory, University of Science and Technology of China, Hefei, Anhui 230088, China}
\affiliation{Shanghai Research Center for Quantum Science and CAS Center for Excellence in Quantum Information and Quantum Physics, University of Science and Technology of China, Shanghai 201315, China}

\author{Youjin Deng}
\email{yjdeng@ustc.edu.cn}
\affiliation{Hefei National Research Center for Physical Sciences at the Microscale and School of Physical Sciences, University of Science and Technology of China, Hefei 230026, China}
\affiliation{Department of Modern Physics, University of Science and Technology of China, Hefei, Anhui 230026, China}
\affiliation{Hefei National Laboratory,
University of Science and Technology of China, Hefei, Anhui 230088, China}

\begin{abstract}
We study the scaling properties of long-range loop-erased random walks (LR-LERW), where the underlying random walker performs L\'evy-flight-like jumps with a power-law step-length distribution $P(\mathbf{r})\sim |\mathbf{r}|^{-(d+\sigma)}$.
{Using extensive Monte Carlo simulations, we measure the scaling relation $N \sim R^{d_N}$ between the loop-erased step number $N$ and the spatial extent $R$, and determine the geometric exponent $d_N$ for various values of $\sigma$ in spatial dimensions $d = 1, 2,$ and $3$, as well as at the marginal point $\sigma = 2$ in $d=4$ and $5$.}
We observe a continuous crossover from long-range (LR) to short-range (SR) behavior as $\sigma$ increases.
Below the upper critical dimension $d<d_c=4$, for $\sigma < d/2$, loop erasure is asymptotically irrelevant and $d_N=\sigma$, consistent with L\'evy-flight scaling.
For $d/2 < \sigma < 2$, loop erasure becomes relevant and $d_N$ varies continuously toward the SR-LERW value.
{At the marginal points with $\sigma=d/2$ or $\sigma=2$, clear logarithmic corrections are observed.}
At and above the upper critical dimension, $d \geq 4$, the scaling at $\sigma=2$ is found to be $N \sim R^2/\ln R$, consistent with that of the corresponding L\'evy flight.
Our results provide a systematic numerical determination of $d_N(\sigma)$ {for the LR-LERW across dimensions, and} are consistent with $\sigma_* = 2$ as the boundary between LR and SR critical behaviors recently established in a broad variety of statistical models.
\end{abstract}

	
\maketitle


\section{Introduction}
\label{sec:intro}
The loop-erased random walk (LERW), originally introduced by Lawler~\cite{lawler1980self}, is constructed by performing a simple random walk and erasing loops chronologically as they form. 
{Despite its simple definition,} this procedure produces a self-avoiding path with nontrivial statistical properties. 
Over the past decades, LERW {has been widely studied} in probability theory and statistical mechanics. 
It is closely related to conformal field theory (CFT), Schramm--Loewner evolution (SLE), and uniform spanning trees (UST)~\cite{arista2019loop, PhysRevE.89.062101, lawler2011conformal}.
\edit{In particular, the unique path connecting two vertices in a UST has exactly the same distribution as a LERW, a correspondence that underlies Wilson's efficient algorithm for sampling USTs~\cite{wilson1996generating} and links the scaling limits of the two objects.}

\begin{table*}
    \centering
    \small
    \caption{Estimates of the geometric exponent $d_N$, obtained from the power-law fits in Eq.~(\ref{eq:fss}), for various values of $\sigma$ in 1D, 2D, and 3D, with `NN' denoting the nearest-neighbor case for reference.}
    \begin{threeparttable}
    \setlength{\tabcolsep}{4pt}
    \begin{tabular}{cl|cl||cl|cl||cl|cl}
        \hline\hline
        \multicolumn{4}{c||}{1D} & \multicolumn{4}{c||}{2D} & \multicolumn{4}{c}{3D} \\
        \hline
        $\sigma$ & ~~~~$d_N$ & $\sigma$ & ~~~$d_N$
        & $\sigma$ & ~~~~$d_N$ & $\sigma$ & ~~~$d_N$
        & $\sigma$ & ~~~~$d_N$ & $\sigma$ & ~~~$d_N$ \\
        \hline
        0.1 & 0.1000(1) & 1.2 & 0.585(2)
        & 0.1 & 0.1000(1) & 1.2 & 1.044(2)
        & 0.1 & 0.1001(1) & 1.2 & 1.199(1) \\
        
        0.2 & 0.2002(5) & 1.3 & 0.601(2)
        & 0.2 & 0.2002(2) & 1.3 & 1.063(2)
        & 0.2 & 0.2000(1) & 1.3 & 1.300(1) \\
        
        0.3 & 0.300(1) & 1.4 & 0.625(2)
        & 0.3 & 0.3001(1) & 1.4 & 1.082(3)
        & 0.3 & 0.3000(1) & 1.4 & 1.396(3) \\
        
        0.4 & 0.396(5) & 1.5 & 0.656(1)
        & 0.4 & 0.4001(1) & 1.5 & 1.098(3)
        & 0.4 & 0.4000(1) & 1.5 & 1.48(1) \\
        
        0.5 & 0.472(3) & 1.6 & 0.698(2)
        & 0.5 & 0.5002(2) & 1.6 & 1.117(4)
        & 0.5 & 0.5000(1) & 1.6 & 1.543(3) \\
        
        0.6 & 0.508(3) & 1.7 & 0.749(4)
        & 0.6 & 0.6001(1) & 1.7 & 1.141(2)
        & 0.6 & 0.6000(1) & 1.7 & 1.579(2) \\
        
        0.7 & 0.530(2) & 1.8 & 0.835(6)
        & 0.7 & 0.700(1) & 1.8 & 1.163(4)
        & 0.7 & 0.7002(2) & 1.8 & 1.60(1) \\
        
        0.8 & 0.541(2) & 1.9 & 0.95(4)
        & 0.8 & 0.801(1) & 1.9 & 1.21(2)
        & 0.8 & 0.8000(1) & 1.9 & 1.62(1) \\
        
        0.9 & 0.551(2) & 2.0 & 0.98(2)
        & 0.9 & 0.899(1) & 2.0 & 1.24(1)
        & 0.9 & 0.8999(2) & 2.0 & 1.627(10) \\
        
        1.0 & 0.560(1) & 2.5 & 1.002(3)
        & 1.0 & 0.98(1) & 2.5 & 1.251(1)
        & 1.0 & 1.0001(2) & 2.5 & 1.625(2) \\
        
        1.1 & 0.571(2) & NN  &  1
        & 1.1 & 1.021(3) & NN & 5/4
        & 1.1 & 1.099(1) & NN & 1.62400(5)~\cite{PhysRevE.82.062102} \\
        \hline\hline
    \end{tabular}
    \label{tab:dn_all}
    \end{threeparttable}
\end{table*}

\begin{table}[t]
    \centering
    \small
    \caption{Estimates of the geometric exponent $d_N$ and the logarithmic correction exponent $\hat d_N$, obtained using Eq.~(\ref{eq:log}), at the marginal points $\sigma=d/2$ and $\sigma=2$ in 1D, 2D, and 3D. In 3D at $\sigma=2$, no stable logarithmic fit could be identified within numerical resolution. For 4D and 5D, we did not perform explicit fits, but the logarithmic behavior can be seen clearly in Fig.~\ref{fig:log_s_2_45}.}
    \begin{threeparttable}
    \setlength{\tabcolsep}{4.5pt}
    \begin{tabular}{c c l l}
        \hline\hline
        $d$ & $\sigma$ & $~~~d_N$ & $\hat d_N$ \\
        \hline
        1 & 0.5 & 0.498(3) & -0.40(1) \\
          & 2.0 & 1.01(2) & -0.80(2) \\ \hline
        2 & 1.0 & 1.01(1) & -0.37(2) \\
          & 2.0 & 1.249(1) & -0.40(3) \\ \hline
        3 & 1.5 & 1.499(2) & -0.23(2) \\
          & 2.0 & 1.627(10) & --- \\ \hline
        4 & 2.0 & 2         &  -1  \\
          & NN  & 2         & -1/3  \\ \hline
        5 & 2.0 & 2         & -1    \\
        \hline\hline
    \end{tabular}
    \label{tab:dn_log}
    \end{threeparttable}
\end{table}

\begin{figure}[ht]
  \centering
  \includegraphics[width=\linewidth]{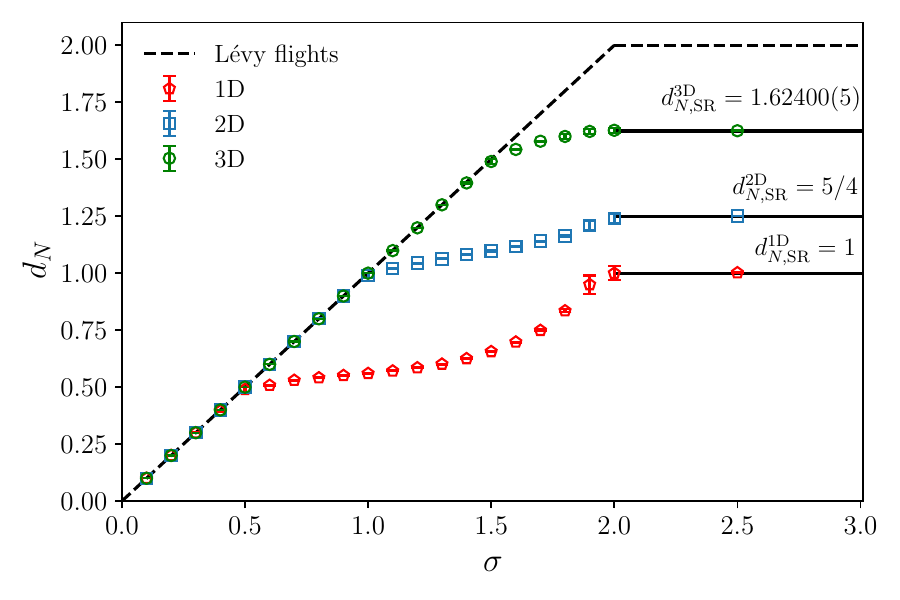}
  \vspace*{-8mm}
    \caption{Geometric exponent $d_N$ as a function of $\sigma$ in 1D (red), 2D (blue), and 3D (green). 
    The estimates are obtained from finite-size scaling fits using Eq.~(\ref{eq:fss}) and Eq.~(\ref{eq:log}).
    The black dashed line shows the L\'evy-flight reference behavior, $d_N=\sigma$ for $\sigma<2$ and $d_N=2$ for $\sigma>2$.
    The black solid lines mark the SR-LERW limits $d_{N,\text{SR}}^{\mathrm{1D}}=1$, $d_{N,\text{SR}}^{\rm{2D}}=5/4$, and $d_{N,\text{SR}}^{\rm{3D}}=1.62400(5)$~\cite{PhysRevE.82.062102}.}
  \label{fig:df_vs_sigma}
\end{figure}

Most established analytical and numerical results concern the short-range (SR) case, typically defined on nearest-neighbor (NN) lattices. In one dimension (1D), the walk is trivially self-avoiding, resulting in a straight trajectory with $N=R$, where $N$ is the loop-erased step number and $R$ is the spatial extent.
{However, the behavior becomes more complex in 2D}: LERW is exactly described by SLE with parameter 2 (SLE$_2$)~\cite{schramm2000scaling,lawler2011conformal}, corresponding to a conformally invariant curve that scales as $N \sim R^{d_N}$ with geometric exponent $d_N=5/4$.
In 3D, extensive numerical simulations have shown that LERW remains a nontrivial geometric object, with geometric exponent $d_N = 1.62400(5)$~\cite{PhysRevE.82.062102}. 
Although no exact analytical solution is available in 3D, recent perturbative field-theoretic approaches~\cite{wiese2019field, PhysRevLett.123.197601} have substantially improved the theoretical understanding of the problem. 
By relating LERW to the O$(n)$ $\phi^4$ theory at $n=-2$ and to the depinning field theory of charge-density waves, these works yielded the estimate $d_N = 1.6243(1)$, in excellent agreement with the numerical result.
Finally, at and above the upper critical dimension $d_c=4$, the system is expected to enter a mean-field (MF) regime in which loop erasure becomes asymptotically irrelevant, and the geometry is therefore governed by Gaussian scaling, with $d_N=2$.
Specifically, at the upper critical dimension $d=d_c$, the MF scaling acquires logarithmic corrections.
The scaling relation is modified to $N \sim R^{2} (\ln R)^{-1/3}$, rather than a pure power law~\cite{wu2019four,grassberger2009scaling}.

In many physical systems, including epidemic spreading and transport in complex media~\cite{ZHAO2023243,10.1093/mnras/staf1474}, motion is not purely diffusive but involves rare, arbitrarily long jumps, as described by L\'evy-flight statistics~\cite{bouchaud1990anomalous, chechkin2008introduction}. 
In such processes, the step distribution decays algebraically as
\begin{align}
P(\mathbf{r}) \propto |\mathbf{r}|^{-(d+\sigma)},
\end{align}
where $\mathbf{r}$ is the displacement vector of the step and the exponent $\sigma$ controls the effective range of motion. 
For  L\'evy flights~\cite{bouchaud1990anomalous}, the scaling relation exhibits a crossover from $N \sim R^{\sigma}$ for $\sigma < 2$ to the normal diffusive behavior $N \sim R^{2}$ for $\sigma > 2$. The case $\sigma = 2$ is marginal and is characterized by a logarithmic correction, $N \sim R^{2} (\ln R)^{-1}$, marking the boundary between LR and SR diffusion.

A natural question is therefore how the scaling behavior of LERW is modified when the underlying random walk follows L\'evy-flight statistics rather than SR dynamics.
LR jumps modify the return probability~\cite{GillisWeiss1970} and thus the frequency of self-intersections, directly affecting the relevance of loop erasure. 
{The insights from L\'evy-flight and related statistical models~\cite{xiao2025universality} suggests the following scaling behavior for LR-LERW.}
For $d<4$ and $\sigma < d/2$, the walker makes sufficiently long jumps that revisits are strongly suppressed, rendering loop erasure asymptotically irrelevant and yielding $d_N = \sigma$, identical to L\'evy flights. 
{In contrast, in} the intermediate regime $d/2 < \sigma < \sigma_*$, self-intersections occur with finite probability and loop erasure becomes relevant, leading to a nontrivial function $d_N(\sigma)$ that interpolates between $d/2$ and the SR value. {Here, $\sigma_*$ denotes the boundary between LR and SR universality.}
For $\sigma > \sigma_*$, LR hopping {becomes} irrelevant and the system should recover the SR universality class. 
{Furthermore, both} marginal points $\sigma = d/2$ and $\sigma = \sigma_*$ may exhibit logarithmic corrections. 
At $d = 4$, one expects $d_N = \sigma$ for $\sigma < 2$, {whereas} short-range-type logarithmic scaling for $\sigma > 2$, of the form $N \sim R^{2} (\ln R)^{-1/3}$~\cite{wu2019four,grassberger2009scaling}. {However, the scaling behavior at the LR marginal point $\sigma = 2$ remains an open question.} If one writes the correction form at this point as $ N \sim R^{2}(\ln R)^{\hat d_N}, $ then one {would expect} $ \hat d_N \le -1.$ This follows because, for a fixed spatial extent $R$, loop erasure can only reduce the number of retained steps, so the loop-erased path cannot be longer than the underlying marginal {L\'evy flight.}
{Finally, for} $d > 4$, the generic scaling is expected to {be the same as} that of L\'evy flights. 


In this work, we introduce and systematically investigate the LR-LERW by implementing a power-law step distribution controlled by $\sigma$ and measuring the scaling relation $N \sim R^{d_N(\sigma)}$.
Our results reveal a unified three-regime structure, as summarized in Fig.~\ref{fig:df_vs_sigma} {for} $d=1,2,3$: for $\sigma<d/2$, the numerical results indicate $d_N = \sigma$, {consistent with L\'evy-flight scaling}; 
for $d/2<\sigma<2$, a crossover regime emerges in which loop erasure progressively reshapes the geometry, {leading to} a nontrivial, continuously varying $d_N(\sigma)$; 
for $\sigma>2$, the system flows to the SR-LERW universality class, recovering the known SR geometric exponents in each spatial dimension. The {resulting} geometric exponents for $d=1,2,3$ are summarized in Table~\ref{tab:dn_all} and Table~\ref{tab:dn_log}.
In addition, at $\sigma = d/2$, logarithmic corrections are observed, as shown in Fig.~\ref{fig:log_s_d_2}, signaling marginal behavior. At $\sigma=2$, logarithmic corrections are also expected at the crossover boundary, as illustrated in Fig.~\ref{fig:log_s_2}, although in $d=3$ they are very weak and might be absent within numerical resolution.
Moreover, for $d=4$ and $5$, our simulations at $\sigma=2$ {suggest} a logarithmic correction exponent $\hat d_N=-1$, as shown in Fig.~\ref{fig:log_s_2_45}.
{Taken together, our results establish $\sigma_*=2$ as the boundary separating the LR and SR critical behaviors for the LR-LERW, independent of the spatial dimensionality.}
Our study therefore provides a systematic numerical {characterization} of $d_N(\sigma)$ of LR-LERW in $d=1,2,3$, {alongside} evidence for the $\hat d_N=-1$ logarithmic correction at $\sigma=2$ in $d=4$ and $5$. {These results} offer quantitative benchmarks and guidance {for future investigations.}

Related studies of other LR systems, including the O$(n)$ model and percolation with power-law interactions, have revealed analogous crossover structures; our results are consistent with the $\sigma_*=2$ scenario reported in Refs~\cite{fisher1972, picco2012, blanchard2013, xiao2025two, yao2025nonclassical, liu2025two, xiao2025universality, yao2025spontaneous, xiao2025sak, li20264}.

The remainder of this paper is organized as follows. Section~\ref{sec:sim} provides a brief overview of the simulation method. Section~\ref{sec:res} presents the main numerical results\edit{, Section~\ref{sec:insights} provides an argument for the marginal point $\sigma=2$}, and Section~\ref{sec:sum} concludes the paper with a summary of our findings.


\section{Simulation}
\label{sec:sim}

We consider the LR-LERW on a $d$-dimensional infinite hypercubic lattice. 
The walk starts from the origin and proceeds according to a L\'evy-type jump distribution~\cite{bouchaud1990anomalous}. 
At each step, a random displacement vector $\mathbf{r}$ is drawn with probability proportional to $|\mathbf{r}|^{-(d+\sigma)}$, where $\sigma>0$ controls the strength of LR hopping.

In $d$ dimensions, the volume element in spherical coordinates is $r^{d-1} dr\, d\Omega$, where $d\Omega$ denotes the angular element. 
Since the distribution is isotropic, it is convenient to separate the radial and angular parts. 
The radial probability density is therefore proportional to
\begin{align}
r^{-(d+\sigma)} r^{d-1} dr = r^{-(1+\sigma)} dr. \nonumber
\end{align}
Because this expression diverges as $r \to 0$, we introduce a lower cutoff $r_0$ on the step length to ensure normalizability. 
The properly normalized cumulative distribution function is then
\begin{align}
F(r) 
= \frac{\int_{r_0}^{r} r'^{-(1+\sigma)} dr'}{\int_{r_0}^{\infty} r'^{-(1+\sigma)} dr'}
= 1 - \left(\frac{r}{r_0}\right)^{-\sigma}. \nonumber
\end{align}
Using inverse transform sampling, we draw a uniform random number $u \in (0,1]$ and solve $F(r)=1-u$, which yields the sampled step length
\begin{align}
r = r_0\, u^{-1/\sigma}.
\end{align}
After determining the step length, a random direction is selected uniformly over the $d$-dimensional unit sphere, giving a normalized vector $\hat{\mathbf{e}}$, and the continuous displacement is defined as $\mathbf{r} = r\,\hat{\mathbf{e}}$.

Since the motion takes place on a discrete lattice, we project the continuous displacement $\mathbf{r}$ onto lattice coordinates by rounding each component to the nearest integer,
\begin{align}
r_i = \mathrm{round}(r\,\hat{e}_i), \quad i=1,\dots,d,
\end{align}
and update the position via $\mathbf{x} \leftarrow \mathbf{x} + \mathbf{r}$. 
Due to lattice discretization, we set $r_0 = 1/2$ in 1D and $r_0 = \sqrt{2}/2$ in 2D such that, in the formal limit $\sigma \to \infty$, the dynamics reduces to a NN walk. For 3D and higher dimensions, we simply take $r_0 = 1$. Under the present discretization scheme, no choice of $r_0$ can, in general, make the $\sigma \to \infty$ limit identical to a NN walk in these dimensions. Nevertheless, this microscopic difference does not affect the large-scale scaling properties studied in this work.

Visited lattice sites are stored in a hash table, enabling $\mathcal{O}(1)$ average-time detection of previously visited sites{~\cite{fangLogarithmicFinitesizeScaling2021}}.
Whenever the walker revisits an occupied site, the closed loop is identified and erased according to the standard LERW rule, and all sites forming the loop are removed from the trajectory.
The walk continues until it exits the simulation domain, defined by the condition that any component of $\mathbf{x}$ exceeds a prescribed cutoff $R$. 
At that point, the total number of steps $N$ in the loop-erased path is recorded. 
To improve sampling efficiency across scales, we successively measure $N$ at increasing cutoff radii $R = 2^3, 2^4, 2^5, \dots, 2^{14}, 2^{15}$. 
The walk proceeds until it exceeds the largest cutoff, so that intermediate $R$ values provide additional measurements across scales without restarting the process.

For $d=1,2,$ and $3$, simulations are carried out for $0.1 \le \sigma \le 2.0$ with increments of $0.1$, supplemented by additional points at $\sigma = 2.5$ to verify the recovery of SR behavior. For $d=4$ and $5$, we focus on the marginal point $\sigma=2$.
All data are averaged over $10^5$ independent realizations. 
The geometric exponent $d_N$ is extracted from the scaling relation $N \sim R^{d_N}$.


\section{Results}
\label{sec:res}

In this section, we present the numerical results for the LR-LERW. Figure~\ref{fig:df_vs_sigma} summarizes the dependence of the geometric exponent $d_N(\sigma)$ on $\sigma$ for $d=1,2,$ and $3$. As $\sigma$ increases, the system undergoes a continuous crossover from a L\'evy-flight-like regime to the SR-LERW regime. The black dashed line denotes the L\'evy-flight reference behavior, with $d_N=\sigma$ for $\sigma<2$ and $d_N=2$ for $\sigma>2$, while the black solid lines indicate the known SR-LERW limit in each dimension.

For $\sigma<d/2$, the data in each dimension agree with the L\'evy-flight scaling within error bars, indicating that loop erasure is asymptotically irrelevant. For $\sigma>d/2$, the estimates deviate from the L\'evy-flight line, showing that self-intersections and loop erasure become relevant. Once $\sigma>2$, the curves approach the SR-LERW plateaus, indicating a crossover to the SR universality class. Moreover, {at} the two boundaries, $\sigma=d/2$ and $\sigma=2$, the data suggest marginal effects and possible logarithmic corrections.


\subsection{Power-Law Scaling}

{We begin by analyzing the power-law scaling for $d=1,2,$ and $3$, with particular emphasis on the fitting procedure and the resulting estimates. The geometric exponent $d_N$ is extracted from the scaling relation between the loop-erased step number $N$ and the spatial extent $R$ of the trajectory. For finite systems, we carry out a systematic finite-size scaling analysis using the fitting form
\begin{align}
N = R^{d_N} \left(a_0 + a_1 R^{-y_1} + a_2 R^{-y_2}\right) + c,
\label{eq:fss}
\end{align}
where $y_1$ and $y_2$ denote correction-to-scaling exponents, and $c$ accounts for a non-singular background contribution. This ansatz captures the leading power-law behavior together with subleading corrections, thereby allowing a controlled and precise determination of $d_N$ over a broad range of $\sigma$.

As a precaution against correction-to-scaling terms that we missed including in the fitting ansatz, we impose a lower cutoff $R \ge R_{\min}$ on the data points admitted in the fit and systematically study the effect on the residuals $\chi^2$ value by increasing $R_{\min}$. In general, the preferred fit for any given ansatz corresponds to the smallest $R_{\min}$ for which the goodness of the fit is reasonable and for which subsequent increases in $R_{\min}$ do not cause the $\chi^2$ value to drop by vastly more than one unit per degree of freedom. In practice, by ``reasonable'' we mean that $\chi^2/\mathrm{DF} \approx 1$, where $\mathrm{DF}$ is the number of degrees of freedom. The systematic error is estimated by comparing estimates from various sensible fitting ansatz.

\subsubsection{One-Dimensional Case}
We begin with the one-dimensional case, where the MF regime is expected for $\sigma<1/2$. In this interval, the data closely follow the L\'evy-flight scaling. For $0.1 \le \sigma \le 0.4$, stable fits to Eq.~\eqref{eq:fss} are obtained from the leading term, supplemented when necessary by the background $c$, and the estimates in Table~\ref{tab:dn_all} are fully consistent with $d_N=\sigma$. As $\sigma$ approaches $1/2$, correction terms become more visible; at $\sigma=0.4$, at least one subleading term is needed.

\begin{figure}[ht]
  \centering
  \includegraphics[width=\linewidth]{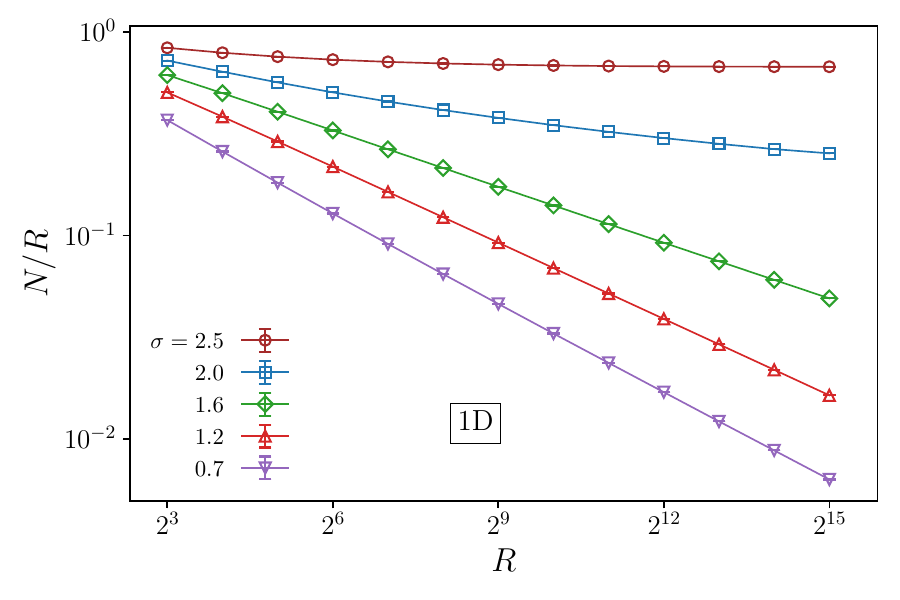}
  \vspace*{-8mm}
    \caption{Log--log plot of $N/R$ versus $R$ in 1D for $\sigma=0.7$, $1.2$, $1.6$, $2.0$, and $2.5$ from bottom to top. The slanted curves for $\sigma<2$ show deviations from the SR scaling. The curve at $\sigma=2.0$ is still clearly non-horizontal, implying a logarithmic correction at the marginal point, whereas the nearly horizontal curve at $\sigma=2.5$ indicates recovery of the SR limit $d_N=1$.}
  \label{fig:1d_pw}
\end{figure}

The regime $0.6\le\sigma\le1.9$ lies in the crossover region, where loop erasure becomes relevant, and $d_N$ deviates from the L\'evy-flight value. Representative fits for $\sigma=0.7$, $1.2$, and $1.6$ are shown in Table~\ref{tab:1d_pw}, and Fig.~\ref{fig:1d_pw} shows the corresponding scaling behavior. The curve for $\sigma=2.5$ is nearly horizontal, indicating that the scaling has already returned to the 1D SR limit $d_N=1$. By contrast, the curves for $\sigma<2$ have clear slopes, showing deviations from SR behavior. In particular, the curve at $\sigma=2$ is still visibly non-horizontal, which suggests a logarithmic correction at this marginal point. For $\sigma=0.7$ and $1.2$, leaving $y_1$ free already gives stable fits with $\chi^2/{\rm DF}$ close to unity. At $\sigma=1.6$, however, $y_1$ cannot be determined reliably, so we repeat the fits with several fixed values of $y_1$ and obtain stable estimates of $d_N$. As $\sigma\to2$, the analysis becomes more delicate, and the uncertainty grows, most notably at $\sigma=1.9$, where we obtain $d_N=0.95(4)$.

\begin{table}[!ht]
    \centering
    \caption{Fitting results for the 1D LR-LERW using the ansatz Eq.~\eqref{eq:fss} with $a_2=0$ and $c=0$.}
    \label{tab:1d_pw}
    \begin{tabular}{l|clllll}
    \hline\hline
        $\sigma$ & $R_{\rm min}$ & ~~~~~$d_N$ & ~~~$a_0$ & ~~~$a_1$ & ~~~$y_1$ & $\chi^2$/DF  \\ \hline
        0.7      & $2^5$           & 0.5323(8)         & 0.809(9) & 0.341(3) & 0.32(1) & 5.2/7  \\ 
        ~        & $2^6$           & 0.532(1)          & 0.82(1)  & 0.346(8) & 0.34(2) & 4.7/6  \\
        ~        & $2^7$           & 0.530(1)          & 0.83(1)  & 0.38(3)  & 0.39(4) & 3.2/5  \\
        ~        & $2^8$           & 0.530(1)          & 0.83(2)  & 0.39(8)  & 0.39(8) & 3.2/4  \\ \hline

        1.2      & $2^4$           & 0.5847(1)	       & 1.234(1) & -0.17(2)  &0.78(6) &  6.8/8  \\ 
        ~        & $2^5$           & 0.5847(1)	       & 1.234(2) & -0.15(5)  &0.7(1) &  6.7/7  \\ 
        ~        & $2^6$           & 0.5842(5)	       & 1.240(8) & -0.06(2)  &0.4(2) &  5.2/6  \\ 
        ~        & $2^7$           & 0.584(2)	       & 1.25(3)  & -0.050(9) &0.3(4) &  5.1/5  \\  \hline
        
        1.6      & $2^8$           & 0.6972(1)	       & 1.146(1) & 1.7(2)   & 1 &  7.2/5  \\ 
        ~        & $2^9$           & 0.6974(1)	       & 1.144(1) & 2.3(4)   & 1 &  4.6/4  \\ 
        ~        & $2^8$           & 0.6974(1)	       & 1.143(1) & 0.65(7)  & 0.8 &  4.8/5 \\ 
        ~        & $2^9$           & 0.6975(1)	       & 1.141(2) & 0.8(1)   & 0.8 &  3.5/4  \\ 
        ~        & $2^8$           & 0.6981(1)	       & 1.134(1) & 0.19(1)  & 0.5 &  2.3/5  \\ 
        ~        & $2^9$           & 0.6981(2)	       & 1.133(2) & 0.20(2)  & 0.5 &  2.2/4  \\  \hline\hline
    \end{tabular}
\end{table}

For $\sigma=2.5$, a free fit of $y_1$ becomes unstable, but fixing $y_1=1$ or $0.5$ yields consistent estimates. The resulting value $d_N=1.002(3)$ agrees with the 1D SR-LERW exponent, confirming SR behavior for $\sigma>2$.

The red symbols in Fig.~\ref{fig:df_vs_sigma} summarize this smooth evolution from the L\'evy-flight branch to the SR limit, and the corresponding estimates are listed in Table~\ref{tab:dn_all}.

\subsubsection{Two-Dimensional Case}
We next turn to two dimensions, where the MF regime is expected for $\sigma<1$. The data again follow this scenario closely. For {$0.1 \le \sigma \le 0.6$}, the leading term alone yields stable fits with $\chi^2/{\rm DF}\approx1$. In the range {$0.7 \le \sigma\le0.9$}, finite-size corrections become visible, and at least one subleading term is needed. The resulting estimates in Table~\ref{tab:dn_all} remain fully consistent with $d_N=\sigma$ throughout the $\sigma<1$ regime.

\begin{figure}[ht]
  \centering
  \includegraphics[width=\linewidth]{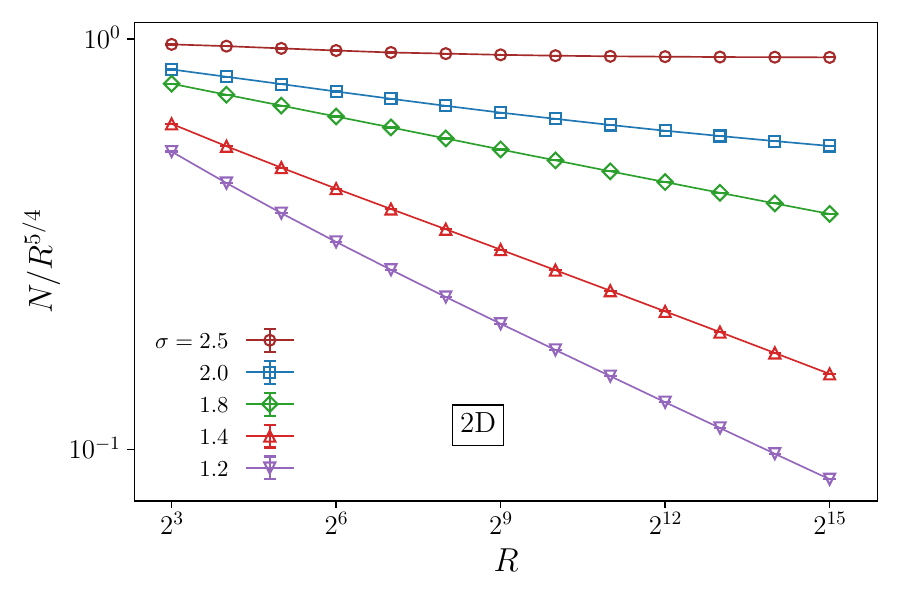}
  \vspace*{-8mm}
    \caption{Log--log plot of $N/R^{5/4}$ versus $R$ in 2D for $\sigma=1.2$, $1.4$, $1.8$, $2.0$, and $2.5$ from bottom to top. The non-horizontal behavior for $\sigma<2$ indicates deviations from the SR scaling. The curve at $\sigma=2.0$ is still visibly non-horizontal, implying a logarithmic correction at the marginal point, whereas the nearly horizontal curve at $\sigma=2.5$ indicates recovery of the SR limit $d_N=5/4$.}
  \label{fig:2d_pw}
\end{figure}

For $1.1\le\sigma\le1.9$, the system lies in the crossover regime between the MF and SR behaviors. Representative fits for $\sigma=1.2$, $1.4$, and $1.8$ are collected in Table~\ref{tab:2d_pw}, and the scaling behavior is shown in Fig.~\ref{fig:2d_pw}. The curve for $\sigma=2.5$ is already nearly horizontal, indicating that the scaling has returned to the 2D SR limit $d_N=5/4$. By contrast, the curves for $\sigma<2$ show clear departures from the SR reference. In particular, the curve at $\sigma=2$ still has a visible slope, again suggesting a logarithmic correction at the marginal point. At $\sigma=1.2$, leaving $y_1$ free gives stable fits. For $\sigma=1.4$ and $1.8$, $y_1$ cannot be determined reliably, so we perform fits with several fixed values. The resulting estimates of $d_N$ remain stable under these choices and under changes of $R_{\min}$. As in 1D, the analysis becomes more delicate near $\sigma=2$.

\begin{table}[!ht]
    \centering
    \caption{Fitting results for the 2D LR-LERW using the ansatz Eq.~\eqref{eq:fss} with $a_2=0$ and $c=0$.}
    \label{tab:2d_pw}
    \begin{tabular}{l|clllll}
    \hline\hline
        $\sigma$ & $R_{\rm min}$ & ~~~~~$d_N$ & ~~~$a_0$ & ~~~$a_1$ & ~~~$y_1$ & $\chi^2$/DF  \\ \hline
        1.2      & $2^3$           & 1.0440(2)         & 0.718(2)  & 0.262(2) & 0.48(1) & 9.2/9  \\ 
        ~        & $2^4$           & 1.0437(3)         & 0.721(2)  & 0.271(6) & 0.50(1) & 7.1/8  \\
        ~        & $2^5$           & 1.0437(4)         & 0.721(3)  & 0.27(1)  & 0.50(2) & 7.1/7  \\ \hline

        1.4      & $2^7$           & 1.0811(2)	       & 0.886(2) & -0.14(1)  &0.5 &  10.0/6  \\ 
        ~        & $2^8$           & 1.0808(3)         & 0.888(2) & -0.16(2)  &0.5 &  7.5/5  \\ 
        ~        & $2^7$           & 1.0797(2)	       & 0.902(2) & -0.096(6) &0.3 &  5.0/6  \\ 
        ~        & $2^8$           & 1.0795(3)	       & 0.904(3) & -0.10(1)  &0.3 &  4.4/5  \\  \hline
        
        1.8      & $2^7$           & 1.1636(2)	       & 0.919(2) & 0.33(7)   & 0.8 &  11.3/6  \\ 
        ~        & $2^8$           & 1.1640(2)	       & 0.915(2) & 0.5(2)   & 0.8 &  5.2/5\\ 
        ~        & $2^7$           & 1.1643(3)	       & 0.912(2) & 0.12(1)  & 0.5 &  6.7/6  \\ 
        ~        & $2^8$           & 1.1648(2)	       & 0.907(2) & 0.12(1)  & 0.5 &  2.9/5  \\  \hline\hline
    \end{tabular}
\end{table}

At $\sigma=2.5$, the fitted exponent reaches $1.251(1)$. Although a free fit of $y_1$ is unstable, fixing $y_1=1$ or $0.5$ yields consistent results. This value is indistinguishable from the exact SR-LERW exponent $5/4$, confirming recovery of the SR universality class for $\sigma>2$.

The blue curve in Fig.~\ref{fig:df_vs_sigma} therefore interpolates smoothly from the L\'evy-flight regime for $\sigma<1$ to the SR-LERW value $5/4$ for $\sigma>2$, and the corresponding estimates are listed in Table~\ref{tab:dn_all}.

\subsubsection{Three-Dimensional Case}
In three dimensions, the MF regime is expected to persist up to $\sigma<3/2$. This is well supported by the data. For $0.1 \le \sigma \le 1.0$, the data show very weak finite-size corrections; the leading scaling form already yields stable fits with $\chi^2/{\rm DF}\approx1$. For $1.1\le\sigma\le1.4$, finite-size corrections become visible, and at least one subleading term must be included, but the estimates in Table~\ref{tab:dn_all} remain consistent with $d_N=\sigma$. Thus, throughout $\sigma<3/2$, loop erasure remains asymptotically irrelevant within numerical resolution.

\begin{figure}[ht]
  \centering
  \includegraphics[width=\linewidth]{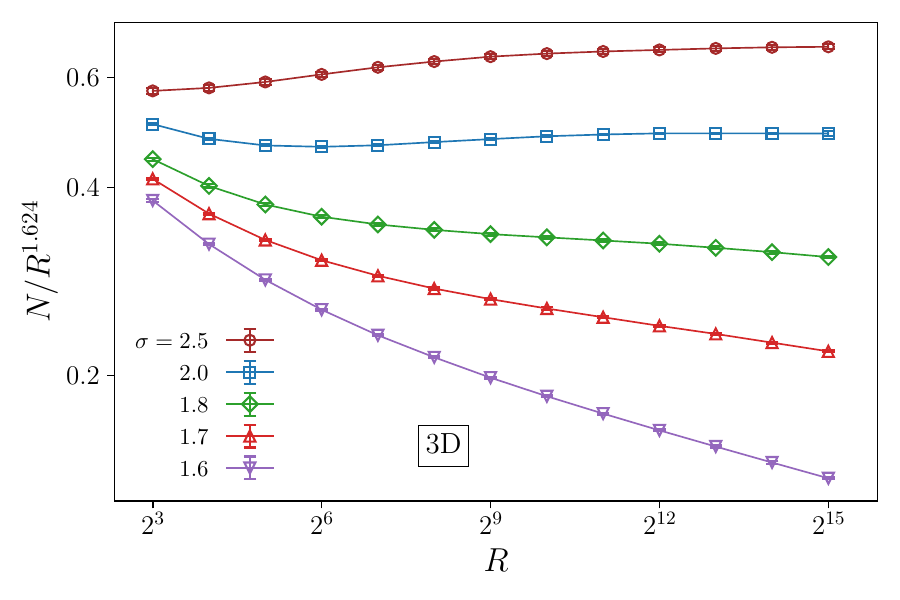}
  \vspace*{-8mm}
    \caption{Log--log plot of $N/R^{1.624}$ versus $R$ in 3D for $\sigma=1.6$, $1.7$, $1.8$, $2.0$, and $2.5$ from bottom to top. The curves for $\sigma<2$ deviate from horizontal behavior. The curves at $\sigma=2.0$ and $2.5$ are both close to horizontal, indicating that any logarithmic correction at $\sigma=2$ is very weak and that the scaling at $\sigma=2.5$ has already returned to the SR limit $d_N \approx 1.624$.}
  \label{fig:3d_pw}
\end{figure}

For $1.6\le\sigma\le1.9$, the system enters the crossover regime and the exponent bends away from the L\'evy-flight line toward the 3D SR-LERW value. Representative fits for $\sigma=1.6$, $1.7$, and $1.8$ are listed in Table~\ref{tab:3d_pw}, and Fig.~\ref{fig:3d_pw} shows the corresponding scaling behavior. The curve for $\sigma=2.5$ is essentially horizontal, indicating recovery of the 3D SR limit. The curves for $\sigma<2$ remain visibly tilted, showing clear deviations from SR scaling. In contrast to 1D and 2D, the curve at $\sigma=2$ is already close to horizontal, suggesting that any logarithmic correction at this marginal point is very weak. At $\sigma=1.6$, leaving $y_1$ free still gives stable fits. At $\sigma=1.7$, $y_1$ becomes less stable, so we compare free and fixed-$y_1$ fits and obtain consistent estimates of $d_N$. For $\sigma=1.8$, reliable fits are obtained only after fixing $y_1$ to representative values such as $1$ or $0.8$. The resulting estimates remain stable under variations of both $R_{\min}$ and $y_1$.

\begin{table}[!ht]
    \centering
    \caption{Fitting results for the 3D LR-LERW using the ansatz Eq.~\eqref{eq:fss} with $a_2=0$ and $c=0$.}
    \label{tab:3d_pw}
    \begin{tabular}{l|clllll}
    \hline\hline
        $\sigma$ & $R_{\rm min}$ & ~~~~~$d_N$ & ~~~$a_0$ & ~~~$a_1$ & ~~~$y_1$ & $\chi^2$/DF  \\ \hline
        1.6      & $2^3$           & 1.5426(4)         & 0.318(1)  & 0.448(4) & 0.584(7) & 13.6/9  \\ 
        ~        & $2^4$           & 1.5430(5)         & 0.316(1) & 0.436(9) & 0.57(1) & 11.3/8  \\
        ~        & $2^5$           & 1.5437(8)         & 0.314(2)  & 0.42(1)  &0.55(1) & 9.4/7  \\
        ~        & $2^6$           & 1.545(1)         & 0.310(3)  & 0.37(2)  & 0.51(3) & 6.4/6  \\ \hline

        1.7      & $2^6$           & 1.5799(5)	       & 0.346(1) & 1.1(3)  &0.95(8) &  7.7/6  \\ 
        ~        & $2^7$           & 1.5793(5)	       & 0.348(1) & 3(2)  &1.2(1) &  5.3/5  \\ 
        ~        & $2^6$           & 1.5796(2)	       & 0.3471(7) & 1.34(3)  &1 &  8.2/7 \\ 
        ~        & $2^7$           & 1.5798(3)	       & 0.3463(9)  & 1.43(7) &1 &  6.4/6  \\  \hline
        
        1.8      & $2^8$           & 1.603(1)	       & 0.386(4) & -0.8(5)   & 1 &  6.6/5  \\ 
        ~        & $2^{9}$        & 1.601(1)	       & 0.393(4) & -2.9(9)   & 1 &  2.8/4  \\ 
        ~        & $2^8$           & 1.603(1)	       & 0.388(4) & -0.3(2)  & 0.8 &  6.1/5 \\ 
        ~        & $2^{9}$        & 1.600(1)	       & 0.396(4) & -1.0(3)   & 0.8 &  2.3/4  \\  \hline\hline
    \end{tabular}
\end{table}

At $\sigma=2.5$, the extracted geometric exponent is $1.625(2)$, in excellent agreement with the known SR-LERW value $1.62400(5)$. This confirms that the crossover to SR behavior is essentially complete once $\sigma>2$.

The green symbols in Fig.~\ref{fig:df_vs_sigma} summarize this crossover from the L\'evy-flight branch for $\sigma<3/2$ to the 3D SR-LERW plateau for $\sigma>2$, and the corresponding estimates are listed in Table~\ref{tab:dn_all}.

\subsection{Logarithmic Corrections at Marginal Points}

{We next examine the marginal points $\sigma = d/2$ and $\sigma = 2$, where the scaling behavior may be modified by logarithmic corrections. We adopt the same fitting procedures as in the previous case, but with an extended fitting ansatz, to account for the possible logarithmic corrections,
\begin{align}
N = R^{d_N} \ln^{\hat{d}_{N}}\left(R/R_0\right)
\left(a_0 + a_1 R^{-y_1} + a_2 R^{-y_2}\right) + c,
\label{eq:log}
\end{align}
where $\hat{d}_{N}$ characterizes the strength of the logarithmic correction and $R_0$ is a microscopic reference scale. In all logarithmic fits, we fix $R_0=1$, as allowing it to vary freely leads to strong parameter correlations without improving the overall fitting stability.}

\subsubsection{Marginal Points at $\sigma = d / 2$}

\begin{figure}[ht]
  \centering
  \includegraphics[width=\linewidth]{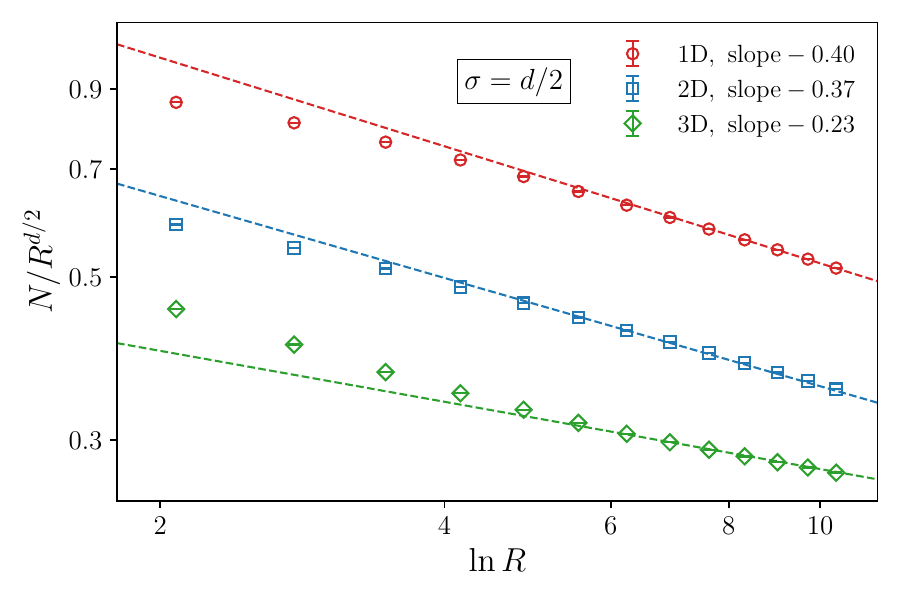}
  \vspace*{-8mm}
    \caption{Log-log plot of $N/R^{d/2}$ versus $\ln R$ at the marginal point $\sigma=d/2$ for 1D, 2D, and 3D. The dashed lines serve as guides to the eye and highlight the logarithmic scaling, with slopes $-0.40(1)$, $-0.37(2)$, and $-0.23(2)$ for 1D, 2D, and 3D, respectively.}
  \label{fig:log_s_d_2}
\end{figure}

At the MF threshold $\sigma=d/2$, pure power-law fits become less stable in all three dimensions, indicating marginal behavior. Figure~\ref{fig:log_s_d_2} shows that, after dividing out the leading factor $R^{d/2}$, the remaining scale dependence is naturally expressed in terms of $\ln R$, suggesting the presence of logarithmic corrections.

In 1D, the marginal point is $\sigma=0.5$. A pure power-law fit gives $d_N=0.472(3)$, but only with a large correction term and $y_1$ close to $0.3$. A fully free logarithmic fit based on Eq.~\eqref{eq:log} is unstable, so we instead fix the leading exponent to $d_N=0.5$ and vary $y_1=1.0$, $0.8$, and $0.5$ to estimate the logarithmic exponent. This gives $\hat d_N=-0.40(1)$, consistent with the slope in Fig.~\ref{fig:log_s_d_2}. Fixing $\hat d_N=-0.40$ then yields $d_N=0.498(3)$ with $\chi^2/{\rm DF}\approx1$. Taking both fitting forms into account, we quote the conservative estimate $d_N=0.49(1)$.

In 2D, at $\sigma=1$, a power-law fit gives $d_N=0.98(1)$ with visible residual drift. Repeating the same procedure, by fixing $d_N=1.0$ and varying $y_1=1.0$, $0.8$, and $0.5$, gives a stable estimate $\hat d_N=-0.37(2)$, again consistent with Fig.~\ref{fig:log_s_d_2}. With $\hat d_N$ fixed, the fit gives $d_N=1.01(1)$ and $\chi^2/{\rm DF}\approx1$. We therefore quote $d_N=0.99(1)$, providing clear evidence for logarithmic corrections at $\sigma=d/2$.

In 3D, the marginal point is $\sigma=1.5$. A pure power-law fit gives $d_N=1.48(1)$ with a relatively large $\chi^2/{\rm DF}$. Applying the same strategy, with $d_N=1.5$ fixed and $y_1=1.0$, $0.8$, and $0.5$, gives $\hat d_N=-0.23(2)$, consistent with the slope shown in Fig.~\ref{fig:log_s_d_2}. Fixing this value yields $d_N=1.499(2)$ with much improved fit quality. Combining both fitting forms, we quote $d_N=1.49(1)$, consistent with MF scaling dressed by a logarithmic correction.

\subsubsection{Marginal Points at $\sigma = 2$}

\begin{figure}[ht]
  \centering
  \includegraphics[width=\linewidth]{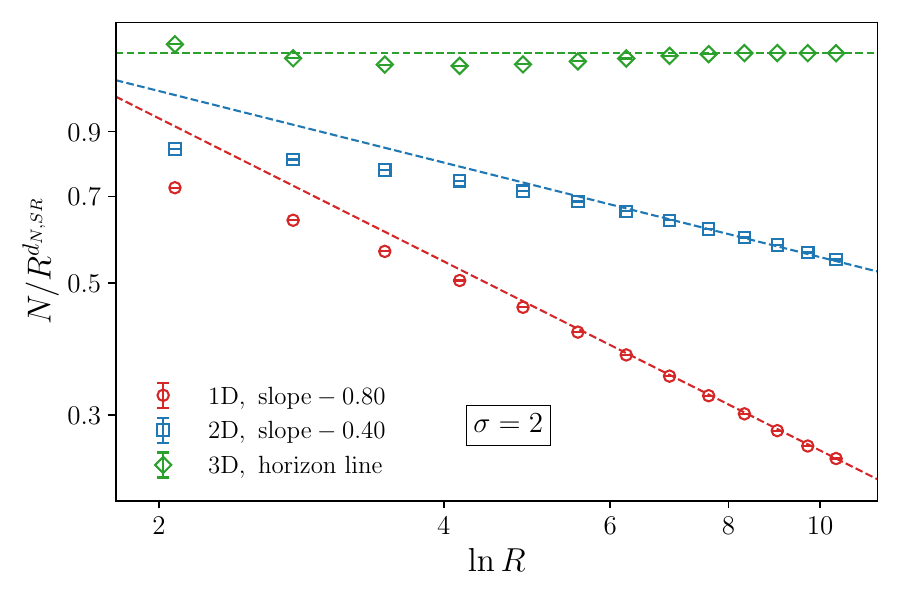}
  \vspace*{-8mm}
    \caption{Log-log plot of $N/R^{d_{N,\rm SR}}$ versus $\ln R$ at $\sigma=2$ for 1D, 2D, and 3D, where $d_{N,\rm SR}=1$ in 1D, $d_{N,\rm SR}=5/4$ in 2D, and $d_{N,\rm SR}=1.62400(5)$~\cite{PhysRevE.82.062102} in 3D. The dashed lines serve as guides to the eye and highlight the logarithmic scaling, with slopes $-0.80(2)$ in 1D and $-0.40(3)$ in 2D. In 3D, the data are nearly horizontal, indicating a very weak or possibly absent logarithmic correction.}
  \label{fig:log_s_2}
\end{figure}

\begin{figure}[ht]
  \centering
  \includegraphics[width=\linewidth]{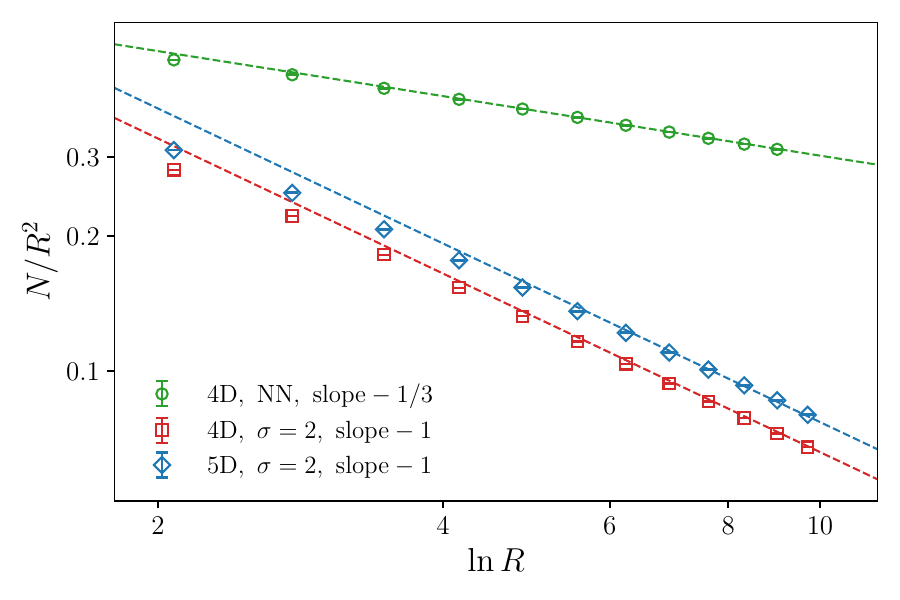}
  \vspace*{-8mm}
    \caption{Log-log plot of $N/R^2$ versus $\ln R$ for the NN LERW in 4D and for the LR-LERW at $\sigma=2$ in 4D and 5D. The dashed lines serve as guides to the eye and highlight the logarithmic scaling, with slopes $-1/3$ for the 4D NN case and $-1$ for the LR cases in both 4D and 5D.}
  \label{fig:log_s_2_45}
\end{figure}

The second marginal point, $\sigma=2$, marks the boundary between the LR and SR regimes. We apply the same analysis below, dividing out the leading factor $R^{d_{N,\rm SR}}$, where $d_{N,\rm SR}=1$ in 1D, $d_{N,\rm SR}=5/4$ in 2D, and $d_{N,\rm SR}=1.62400(5)$~\cite{PhysRevE.82.062102} in 3D. For 4D and 5D, we divide by $R^2$.

In 1D, a pure power-law fit at $\sigma=2$ gives $d_N=0.98(2)$, but with a large correction term, $y_1$ close to $0.25$, and poor fit quality. A fully free logarithmic fit is again unstable, so we fix $d_N=1.0$ and vary $y_1=1.0$, $0.8$, and $0.5$. This gives $\hat d_N=-0.80(2)$, {asymptotically} consistent with the slope in Fig.~\ref{fig:log_s_2}. With $\hat d_N=-0.8$ fixed, the fit gives $d_N=1.01(2)$. Taking both fitting forms into account, we quote $d_N=1.00(3)$.

In 2D, the same pattern persists. A power-law fit gives $d_N=1.24(1)$, again with a large correction term and $y_1=0.14$. Fixing $d_N=1.25$ and varying $y_1=1.0$, $0.8$, and $0.5$ gives $\hat d_N=-0.40(3)$, in agreement with Fig.~\ref{fig:log_s_2}. Fixing this value yields $d_N=1.249(1)$ with $\chi^2/{\rm DF}\approx1$, fully consistent with the SR value $1.25$. Combining both fitting forms, we quote $d_N=1.24(2)$.

The situation at $\sigma=2$ in 3D is more subtle. Our power-law estimate $d_N=1.627(10)$ is already very close to the SR value $1.62400(5)$, but unlike in 1D and 2D, no stable logarithmic fit can be isolated under reasonable variations of the fitting strategy. The data in Fig.~\ref{fig:log_s_2} are nearly horizontal, suggesting that any logarithmic correction, {if present,} is much weaker and remains numerically inconclusive in this case.


Moreover, for $d=4$ and $5$, we focus on the marginal point $\sigma=2$. For $d>4$, the scaling is expected to coincide with that of L\'evy flights, and our 5D data at $\sigma=2$, as shown by the blue line in Fig.~\ref{fig:log_s_2_45}, indicate a slope of $-1$, corresponding to the scaling $N \sim R^2 (\ln R)^{-1}$, namely $\hat d_N=-1$. In 4D, the regime $\sigma<2$ is expected to follow the L\'evy-flight scaling $N \sim R^{\sigma}$, whereas the regime $\sigma>2$ is expected to exhibit a logarithmic correction with $\hat d_N=-1/3$, as shown by the green line in Fig.~\ref{fig:log_s_2_45}. Hence, the case $\sigma=2$ is of particular interest as the marginal LR point. Interestingly, our 4D data again support $N \sim R^2 (\ln R)^{-1}$, as shown by the red line in Fig.~\ref{fig:log_s_2_45}, just as in the 5D case at $\sigma=2$. This suggests that at $(d,\sigma)=(4,2)$ the system has likely already crossed over to L\'evy-flight-like behavior, so that the observed logarithmic correction can be understood as the intrinsic marginal correction of L\'evy flights at $\sigma=2$. \edit{An argument supporting this picture is provided in Sec.~\ref{sec:insights}.}


\edit{
\section{Insights from random-walk intersections at $\sigma=2$}
\label{sec:insights}

The numerical results above show that, at $(d,\sigma)=(4,2)$, the LR-LERW follows the same scaling $N\sim R^2(\ln R)^{-1}$ as the underlying L\'evy flight, whereas the NN LERW at $d=4$ exhibits the distinct logarithmic correction $N\sim R^2(\ln R)^{-1/3}$. We now provide an argument, based on the intersection properties of random walks, to account for this observation.

At the marginal value $\sigma=2$, the second moment of the step-length distribution diverges logarithmically, and the typical displacement of a L\'evy flight after $N$ steps acquires a logarithmic correction,
\begin{align}
R_N \sim \sqrt{N \ln N},
\label{eq:rn}
\end{align}
in contrast to the purely diffusive behavior $R_N\sim N^{1/2}$ for $\sigma>2$. The volume typically explored by such a walk is therefore
\begin{align}
V_N \sim R_N^d \sim N^{d/2} (\ln N)^{d/2}.
\end{align}

Consider two independent walks of $N$ steps each. Each of the $N$ visited sites of one walk is visited by the other walk with a probability of order $N/V_N$, so that the expected number of intersections scales as
\begin{align}
N_{\rm int} \sim N \cdot \frac{N}{V_N} \sim N^{2-d/2} (\ln N)^{-d/2}.
\label{eq:nint}
\end{align}
Self-intersections of a single walk are expected to obey the same scaling. Equation~\eqref{eq:nint} shows that, at $\sigma=2$, intersections are asymptotically absent for $d>4$ and, crucially, also vanish at $d=4$,
\begin{align}
N_{\rm int} \sim (\ln N)^{-2} \to 0 \qquad (d=4,\ \sigma=2).
\end{align}
Loop erasure, which acts only on these rare self-intersections, is thus asymptotically irrelevant at $(d,\sigma)=(4,2)$: the upper critical dimension at $\sigma=2$ lies logarithmically below four, which we denote by $d_c=4^-$. The scaling of the LR-LERW then necessarily coincides with that of the underlying walk. Inverting Eq.~\eqref{eq:rn}, up to constants and slowly varying $\ln\ln R$ terms, gives
\begin{align}
N \sim \frac{R^2}{\ln R},
\label{eq:nr2}
\end{align}
i.e., $d_N=2$ with $\hat d_N=-1$, in full agreement with the 4D and 5D data in Fig.~\ref{fig:log_s_2_45}. This also saturates the bound $\hat d_N\le -1$ argued in Sec.~\ref{sec:intro}.

The situation is markedly different for the NN LERW at $d=4$. There $R_N\sim N^{1/2}$, and Eq.~\eqref{eq:nint} yields $N_{\rm int}\sim O(1)$: self-intersections are marginal, loop erasure remains marginally relevant, and the scaling acquires the characteristic correction $N\sim R^2(\ln R)^{-1/3}$~\cite{wu2019four,grassberger2009scaling}. For the $\sigma=2$ L\'evy flight, the additional $\sqrt{\ln N}$ swelling of the trajectory suppresses intersections by an extra factor $(\ln N)^{-2}$, rendering loop erasure fully irrelevant; the only surviving logarithmic correction is then the intrinsic marginal correction of the L\'evy flight itself, $\hat d_N=-1$. Below four dimensions, $N_{\rm int}$ diverges and loop erasure becomes relevant, consistent with the nontrivial crossover behavior observed in $d=1,2,3$ (Sec.~\ref{sec:res}).

A similar marginal structure occurs in LR-O$(n)$ spin models with interactions decaying as $r^{-(d+\sigma)}$. At $\sigma=2$, the long-wavelength Gaussian kernel takes the form $k^2\ln(1/k)$, differing from the SR form $k^2$ only by a marginal logarithmic factor. 
Consequently, in the ordered phase for $n \geq 2$, the correlation function behaves as $C(r)\simeq c_0+a r^{2-d}/\ln r$ for $d>2$, where $C_0>0$ accounts for the long-range order and the algebraic corrections arise from the Goldstone-mode fluctuations. 
In contrast, the SR case gives the conventional Ornstein--Zernike correction $C(r)\simeq c_0+a r^{2-d}$~\cite{xiao2025universality,xiao2025sak,li20264}. Thus, at $\sigma=2$, the leading power-law behavior is already SR-like, while the marginal LR character survives through logarithmic corrections. This closely parallels the LR-LERW at $(d,\sigma)=(4,2)$, where loop erasure becomes asymptotically irrelevant and the leading scaling reduces to that of the underlying walk, with the marginal long-range character likewise reflected in a logarithmic correction.
}


\section{Discussion}
\label{sec:sum}

We perform a systematic numerical investigation of the long-range loop-erased random walk (LR-LERW), focusing on how the geometric exponent $d_N$ of loop-erased trajectories evolves with the exponent $\sigma$ in the presence of power-law distributed LR steps with probability decaying as $r^{-(d+\sigma)}$. Our results reveal a rich crossover structure that continuously connects the L\'evy-flight regime to the short-range loop-erased random walk (SR-LERW) regime.

Above the upper critical dimension, $d>d_c=4$, the scaling is fully consistent with that of L\'evy flights: for $\sigma < 2$, $d_N = \sigma$, while for $\sigma > 2$, $d_N=2$. At the upper critical dimension, $d=d_c$, the same picture still holds at the level of leading power-law scaling, but with logarithmic corrections, characterized by $\hat{d}_N = -1/3$ for $\sigma>2$. 
When $d<d_c$, for $\sigma < d/2$, loop erasure is asymptotically irrelevant, and the scaling behavior is governed by that of L\'evy flights, with $d_N = \sigma$; for $d/2 < \sigma < 2$, loop erasure become relevant, leading to a crossover regime in which $d_N(\sigma)$ varies continuously toward the SR value. For $\sigma > 2$, the system crosses over to the SR-LERW universality class, recovering the corresponding SR geometric exponent in each spatial dimension.

{Interestingly, the $d_N(\sigma)$ behavior exhibits a striking contrast between 1D and 3D. While the 1D curve initially becomes concave once $\sigma$ exceeds $d/2$ and subsequently displays a sharp convex upturn near $\sigma=2$, the 3D trajectory remains smoothly concave toward the SR limit. This contrast qualitatively reflects the enhanced role of fluctuations in lower spatial dimensions.}

{A central finding of this study is that the LR-SR crossover of the LR-LERW occurs at $\sigma_* = 2$, irrespective of dimensionality. Moreover, at $\sigma=2$}, clear logarithmic corrections are observed {for} $d=1,2,4,$ and $5$, while {for} $d=3$, any such correction appears to be very weak and remains numerically inconclusive. In particular, for $d=4$ and $5$ at $\sigma=2$, our data are consistent with the {standard logarithmic correction form of the L\'evy-flight, namely $N \sim R^2 (\ln R)^{-1}$}. {This suggests that at $(d, \sigma)=(4, 2)$, the system has likely already entered the L\'evy-flight regime, with the logarithmic correction manifesting as the intrinsic marginal signature of L\'evy flights at $\sigma=2$.}

{More broadly, the present results provide another example supporting the $\sigma_*=2$ threshold for LR systems, in line with recent studies across various universality classes~\cite{xiao2025two, yao2025nonclassical, liu2025two, xiao2025universality, yao2025spontaneous, xiao2025sak, li20264}. 
This picture is intrinsically connected to the behavior of L\'evy flights~\cite{xiao2025universality} where $\sigma = 2$ marks the point at which the second moment of the step-length distribution diverges. Regardless of spatial dimensionality, the walk is asymptotically diffusive for $\sigma>2$, where the LR tail is irrelevant at large scales. In contrast, for $\sigma\le 2$, the motion is governed by L\'evy-stable statistics, and the walk becomes super-diffusive, leading to an LR-interaction-dominant regime.}

\edit{A comparison with critical percolation provides a broader perspective on the role of geometric constraints. For the SR-LERW, the upper critical dimension $d_c=4$ is naturally associated with the intersection properties of random-walk trajectories, whereas critical percolation, with its branched and multiply connected cluster geometry, has the higher upper critical dimension $d_c=6$. In the latter case, Milovanov related the persistence of topological constraints below six dimensions to a broader topological and homotopic framework based on the Whitney embedding theorem~\cite{milovanov1997topological}. Although the LERW does not admit a direct mapping onto this framework, both systems illustrate a common qualitative theme: nontrivial critical geometry persists as long as the underlying geometric or topological constraints remain relevant, whereas mean-field behavior emerges once such constraints become asymptotically irrelevant.}

\acknowledgments
We acknowledge the support by the National Natural Science Foundation of China (NSFC) under Grant No. 12204173 and No. 12275263, as well as Quantum Science and Technology-National Science and Technology Major Project (under Grant No. 2021ZD0301900). YD is also supported by the Natural Science Foundation of Fujian Province 802 of China (Grant No. 2023J02032). ZF is also supported by the National Natural Science Foundation of China (NSFC) under Grant No. 12504265.

{\section*{data availability}
The data that support the findings of this article are openly available~\cite{tensofermi2026data}.
}

\appendix
\edit{
\section{Stability of the fitting procedure}
\label{app:fit}

In this appendix, we describe in more detail how the stability of the fitting ansatzes in Eqs.~(\ref{eq:fss}) and (\ref{eq:log}) is assessed numerically, and present representative stability tests.

For a given ansatz, we impose a lower cutoff $R\ge R_{\min}$ on the data and systematically increase $R_{\min}$. The preferred fit corresponds to the smallest $R_{\min}$ for which the goodness of fit becomes reasonable, i.e., $\chi^2/\mathrm{DF}\approx1$, and for which further increases of $R_{\min}$ do not cause the $\chi^2$ value to drop by vastly more than one unit per degree of freedom. A fit is diagnosed as unstable when the freely fitted correction exponent $y_1$ exhibits a pronounced drift or a strongly inflated uncertainty under variations of $R_{\min}$, while a broad range of $y_1$ values yields statistically comparable values of $\chi^2/\mathrm{DF}$. In this situation, $y_1$ is not meaningfully constrained by the data, and we instead perform a series of fits with $y_1$ fixed at representative values, typically $y_1=1.0$, $0.8$, and $0.5$. The spread among the resulting estimates of the leading exponent is then incorporated into the quoted systematic uncertainty.

As a representative example for the power-law ansatz~(\ref{eq:fss}), consider the 1D data at $\sigma=1.6$ listed in Table~\ref{tab:1d_pw}. A free fit of $y_1$ is unstable in this case. With $y_1$ fixed at $1.0$, $0.8$, and $0.5$, the fits yield $d_N=0.6972$--$0.6981$ with acceptable $\chi^2/\mathrm{DF}$: the leading exponent varies only at the $10^{-3}$ level, whereas the correction amplitude $a_1$ changes by an order of magnitude. The leading exponent $d_N$ is thus considerably more stable than the correction terms, and the small spread is reflected in the quoted uncertainty. By contrast, at $\sigma=0.7$ and $1.2$, free fits already converge with stable $y_1$ and $\chi^2/\mathrm{DF}$ close to unity (Table~\ref{tab:1d_pw}), so that no fixing procedure is required.

For the logarithmic ansatz~(\ref{eq:log}), the reference scale is fixed to $R_0=1$ in all fits, since a freely varying $R_0$ is strongly correlated with $\hat d_N$ and the amplitudes without improving the overall fit quality. As a representative example, consider the 1D data at the marginal point $\sigma=2$. A fully free fit is again unstable. In a first step, we fix $d_N=1$ and vary $y_1=1.0$, $0.8$, and $0.5$, which yields a stable logarithmic exponent $\hat d_N=-0.80(2)$, consistent with the slope shown in Fig.~\ref{fig:log_s_2}. In a second step, fixing $\hat d_N=-0.8$ gives $d_N=1.01(2)$ with $\chi^2/\mathrm{DF}\approx1$. The same two-step strategy is applied at the other marginal points, $\sigma=d/2$ and $\sigma=2$, in $d=1,2,$ and $3$ (Sec.~\ref{sec:res}).
}

\end{document}